\documentclass[reprint,superscriptaddress,amsmath,amssymb,aps,prb,floatfix]{revtex4-2}
\usepackage{siunitx}
\usepackage{makecell}
\usepackage[colorlinks,linkcolor=black,anchorcolor=blue, urlcolor=blue,citecolor=blue,unicode]{hyperref}
\usepackage{multirow}
\usepackage{subfigure}
\usepackage{color}
\usepackage[normalem]{ulem}
\usepackage{amsmath}
\usepackage{booktabs,siunitx,array}
\usepackage{booktabs}
\usepackage{ulem}
\usepackage{listings}
\usepackage{import}
\usepackage{xcolor}
\usepackage{framed}
\usepackage{mdwlist}
\usepackage{diagbox}
\usepackage{bm}
\usepackage{graphicx}
\usepackage{dcolumn}
\usepackage{algorithm}
\usepackage{algpseudocode}
\usepackage{csquotes}
\usepackage{array}
\usepackage{subfigure}
\usepackage{siunitx}
\usepackage{amsmath}
\usepackage{ifthen}
\usepackage{threeparttable}
\usepackage{float}

\usepackage{tabularx}
\usepackage{CJK}
\usepackage{indentfirst}

\begin{document}
	
	\title{Disentangling Anomalous Hall Effect Mechanisms and Extra Symmetry Protection in Altermagnetic Systems}
	\author{Yuansheng Bu}
	\affiliation{Beijing National Laboratory for Condensed Matter Physics and Institute of Physics, Chinese Academy of Sciences, Beijing 100190, China}
	\affiliation{University of Chinese Academy of Sciences, Beijing 100049, China}
	\author{Ziyin Song}
	\affiliation{Beijing National Laboratory for Condensed Matter Physics and Institute of Physics, Chinese Academy of Sciences, Beijing 100190, China}
	\affiliation{University of Chinese Academy of Sciences, Beijing 100049, China}
	\author{Zhong Fang}
	\affiliation{Beijing National Laboratory for Condensed Matter Physics and Institute of Physics, Chinese Academy of Sciences, Beijing 100190, China}
	\affiliation{University of Chinese Academy of Sciences, Beijing 100049, China}
	\author{Quansheng Wu}
	\affiliation{Beijing National Laboratory for Condensed Matter Physics and Institute of Physics, Chinese Academy of Sciences, Beijing 100190, China}
	\affiliation{University of Chinese Academy of Sciences, Beijing 100049, China}
	\author{Hongming Weng}
	\email{hmweng@iphy.ac.cn}
	\affiliation{Beijing National Laboratory for Condensed Matter Physics and Institute of Physics, Chinese Academy of Sciences, Beijing 100190, China}
	\affiliation{University of Chinese Academy of Sciences, Beijing 100049, China}
	\affiliation{Condensed Matter Physics Data Center of Chinese Academy of Sciences, Beijing 100190,  China}
	\newcommand{\XR}[1]{\color{red}{\bf #1 }}
	\newcommand{\XB}[1]{\color{blue}{ #1 }}
	\newcommand{\XY}[1]{\color{yellow}{\bf #1 }}
	\begin{abstract}
	We investigate the evolution of Anomalous Hall Conductivity (AHC) in a coplanar and collinear antiferromagnetic system with varying spin canting angles. A tight-binding model based on three $t_{2g}$-orbitals in a body-centered tetragonal lattice is constructed, where the inclusion of third-nearest neighbor hopping is demonstrated to be essential for capturing the characteristic energy band splitting of altermagnetic materials. By employing a symmetry analysis based on spin space groups and treating spin-orbit coupling (SOC) as a perturbation, we theoretically distinguish and numerically verify two origins of the transverse transport: the conventional anomalous Hall effect (AHE) induced by net magnetization and the Crystal Hall Effect (CHE) arising from specific crystal symmetries. Our results show that the conductivity components driven by these two mechanisms follow distinct trigonometric dependencies on the canting angle. Crucially, we identify a hidden $C_{110}$ rotational symmetry that has been previously overlooked in static magnetic group analyses. By expanding the AHC in terms of spin orientation vectors, we demonstrate that this symmetry acts as a bridge connecting distinct magnetic configurations with different canting angles, thereby strictly protecting the equivalence of orthogonal conductivity components in the collinear system.
    \end{abstract}
	\maketitle
	\section{Introduction}
	The interplay between magnetic ordering and topological transport properties has long been a central focus in condensed matter physics. Conventionally, the AHE was believed to be inextricably linked to a non-vanishing net magnetization \cite{hall1,hall2,RevModPhys.82.1539}, appearing primarily in ferromagnets where time-reversal symmetry is macroscopically broken. The underlying mechanism is typically attributed to the Berry curvature of Bloch electrons in momentum space, induced by SOC \cite{RevModPhys.82.1959}. However, recent theoretical and experimental breakthroughs in the field of \textit{altermagnetism} \cite{PhysRevX.12.031042,PhysRevX.12.040501, FeSb2, feng2022anomalous} have challenged this paradigm. These studies reveal that compensated magnetic systems with zero net magnetization can exhibit a non-relativistic spin splitting in the band structure typically with $d$-wave or $g$-wave symmetry imposed by specific crystal symmetries. Consequently, these altermagnets can host a large, non-vanishing AHE despite the absence of net magnetization, provided that the combined symmetry of time-reversal and lattice translation is broken \cite{betancourt2023spontaneous, gonzalez2021efficient}.
	Understanding the microscopic mechanisms governing these transport phenomena, particularly in realistic lattice models, is crucial for both fundamental physics and potential spintronic applications such as terahertz writing speeds and high-density storage \cite{shao2021spin, bose2022tilted}. Recently, a comprehensive macroscopic classification of the AHE in purely collinear altermagnets has been established by treating the Néel vector as an extrinsic parameter under the constraints of non-magnetic crystallographic point groups \cite{xiao2025anomalous}. While the AHE in purely collinear altermagnets and its associated macroscopic Néel textures are well-established, real-world antiferromagnetic materials often deviate from perfect collinearity due to external fields or Dzyaloshinskii-Moriya interactions (DMI), leading to spin canting. In this work, we investigate the evolution of the AHC in a coplanar antiferromagnetic system, specifically focusing on the dependence of AHC on the spin canting angle. The study of such systems is particularly intriguing because these systems host two distinct origins of the Hall response: the conventional AHE induced by the emergent net magnetization, and the so-called CHE \cite{CHE}, which arises purely from the distinct chemical environments of sublattices and is characteristic of the altermagnetic order. Disentangling these two contributions is vital for interpreting transport experiments in canted antiferromagnets.
	
	To elucidate these mechanisms, we construct a tight-binding model based on a body-centered tetragonal lattice. This lattice structure serves as a minimal model for a large class of rutile-structure antiferromagnets (e.g., RuO$_2$, NiF$_2$) \cite{PhysRevB99184432, bai2022observation, NiF2}. Following the method described in Ref. \cite{PhysRevX.15.031006}, we construct the hopping terms for the three $t_{2g}$ orbitals ($d_{xy}$, $d_{yz}$, and $d_{xz}$) within the tight-binding approximation. Crucially, we extend the hopping terms up to the third-nearest neighbor. We demonstrate that the inclusion of third-nearest neighbor hopping is not merely a quantitative correction but is essential to capture the altermagnetic features \cite{Dagnino2024TheLO, PhysRevLett.132.176702}, specifically the characteristic anisotropic band splitting in the Brillouin zone. This model shows excellent agreement with first-principles calculations of the electronic structure of real materials like NiF$_{2}$, RuO$_{2}$ and MnF$_{2}$\cite{ruo2,mnf2}.
	
	Furthermore, we employ a symmetry analysis based on Spin Space Groups (SSG) \cite{brinkman1966SSG, PhysRevX.12.021016, corticelli2022spin,PhysRevX.14.031039,song_constructions_2025,song2025unified} to interpret the AHC behavior. Unlike traditional Magnetic Space Groups (MSG) which couple spin and lattice rotations rigidly, we treat the energy scale of SOC as a perturbation using SSG analysis. By expanding the AHC to the seventh-order infinitesimal in terms of spin orientation vectors, we theoretically distinguish the contributions from net magnetization and crystal symmetry. We demonstrate that the AHC components driven by these two mechanisms follow distinct trigonometric dependencies ($\sin \eta$ vs. $\cos \eta$) on the canting angle $\eta$. Importantly, our analysis uncovers a hidden symmetry protection mechanism: we clarify that the equivalence of specific orthogonal conductivity components in the collinear system is strictly protected by a two-fold rotation symmetry along the diagonal direction—a symmetry operation that connects distinct magnetic configurations and has been previously overlooked in static symmetry analyses.
	
	\section{Theoretical method}
	\subsection{Model constructing}
\begin{figure}[htbp]
		\centering\includegraphics[width=0.9\columnwidth]{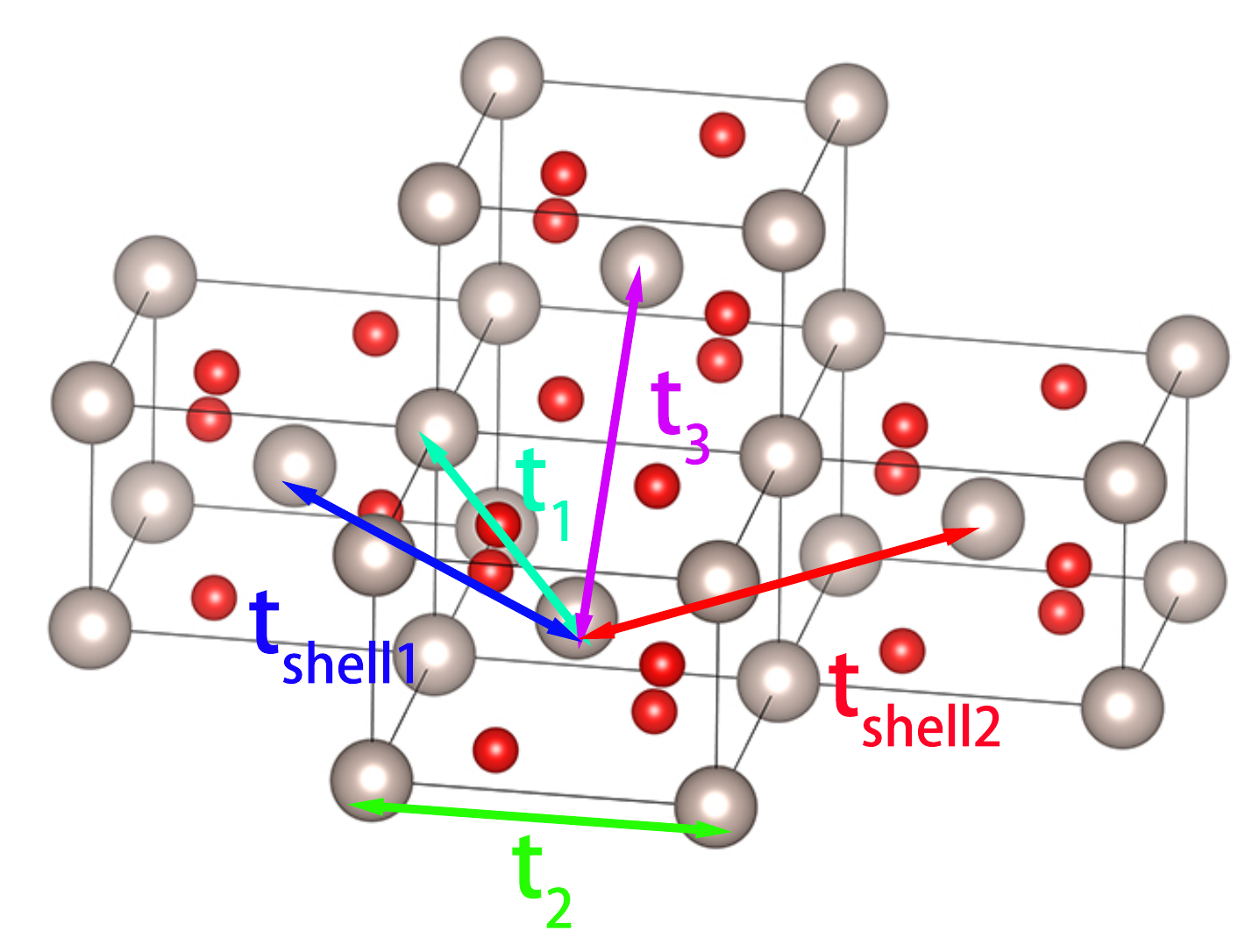}
		\caption{Schematic illustration of the body-centered tetragonal lattice structure adopted in the tight-binding model. The grey and red spheres represent the transition metal (magnetic) and ligand ions, respectively. The colored arrows indicate the hopping integrals considered in the Hamiltonian: $t_1$ (cyan) corresponds to the nearest-neighbor hopping between sublattices, while $t_2$ (green) and $t_3$ (magenta) denote the next-nearest and third-nearest neighbor hoppings, respectively. The vectors $t_{shell1}$ (blue) and $t_{shell2}$ (red) illustrate the anisotropic hopping paths that distinguish the local environments of the two sublattices, which is essential for capturing the altermagnetic symmetry.}
		\label{fig:1}
	\end{figure}

	To investigate the evolution of AHC with respect to the canting angle in this coplanar antiferromagnetic system, we construct a tight-binding model based on three $t_{2g}$ orbitals ($d_{xy}$, $d_{yz}$, $d_{xz}$). A body-centered tetragonal lattice is adopted as the minimal model for a three-dimensional antiferromagnet. As illustrated in Fig. \ref{fig:1}, the structure consists of two sublattices where the body-centered sites (sublattice B) are related to the corner sites (sublattice A) by a four-fold screw axis along the $z$-axis. To capture the characteristic altermagnetic symmetry, the model incorporates hopping terms up to the third-nearest neighbors, parameterized via the Slater-Koster method\cite{PhysRev.94.1498,DAPapaconstantopoulos_2003}. The real-space Hamiltonian is defined as:
	\begin{equation}
		\begin{split}
			H &= \sum_{n=1,2,3} \sum_{a} t_{n}^{a} \sum_{\langle i,j \rangle_{n,a}} c_{i\sigma}^{\dagger}c_{j\sigma} \\
			&\quad + J \sum_{i,\alpha} \sum_{\sigma,\sigma'} \hat{M}_{i\alpha} \left( c_{i\sigma}^{\dagger} \sigma_{\sigma\sigma'}^{\alpha} c_{i\sigma'} \right) \\
			&\quad + \lambda \sum_{i} \sum_{m,m'} \sum_{\sigma,\sigma'} \langle m, \sigma | \boldsymbol{L} \cdot \boldsymbol{S} | m', \sigma' \rangle c_{im\sigma}^{\dagger} c_{im'\sigma'},
		\end{split}
	\end{equation}
	where $t_n^a$ denotes the hopping integral for the $a$-th shell of the $n$-th nearest neighbors, $J$ and $\hat{\boldsymbol{M}_{\alpha}}$ ($\alpha \in \{A, B\}$ denotes the sublattice index) are the coupling strength and direction of magnetization of $\alpha$ sublattice and $\lambda$ is the strength of SOC. Note that while nearest-neighbor $t_1$ and next-nearest-neighbor $t_2$ hoppings maintain a higher symmetry than the original lattice, the third-nearest-neighbor hopping $t_3$ is crucial for lifting certain degeneracies. In a tetragonal lattice, $t_3$ comprises two distinct coordination shells $t_{shell1}$ and $t_{shell2}$ in the $xy$ plane induced by the four-fold screw axis perpendicular to the $xy$ plane ~\cite{Dagnino2024TheLO,PhysRevLett.132.176702} and the equivalent hoppings in the other planes. To explicitly reflect the lattice symmetry, we decompose the Hamiltonian in $k$-space using a sublattice basis:
	\begin{equation}
		H(\boldsymbol{k}) = \begin{pmatrix} H_{AA}(\boldsymbol{k}) & H_{AB}(\boldsymbol{k}) \\ H_{BA}(\boldsymbol{k}) & H_{BB}(\boldsymbol{k}) \end{pmatrix},
	\end{equation}
	where $H_{AA}$ and $H_{BB}$ represent intra-sublattice terms, and $H_{AB}$ accounts for inter-sublattice hopping. Each diagonal block $H_{\alpha\alpha}$ ($\alpha \in \{A, B\}$) is a $6 \times 6$ matrix in the combined orbital and spin space:
	\begin{equation}
		H_{\alpha\alpha} = \begin{pmatrix} H_{\alpha\alpha}^{\uparrow\uparrow} & H_{\alpha\alpha}^{\uparrow\downarrow} \\ H_{\alpha\alpha}^{\downarrow\uparrow} & H_{\alpha\alpha}^{\downarrow\downarrow} \end{pmatrix},
	\end{equation}
	\begin{figure*}[htbp]
		\centering\includegraphics[width=2\columnwidth]{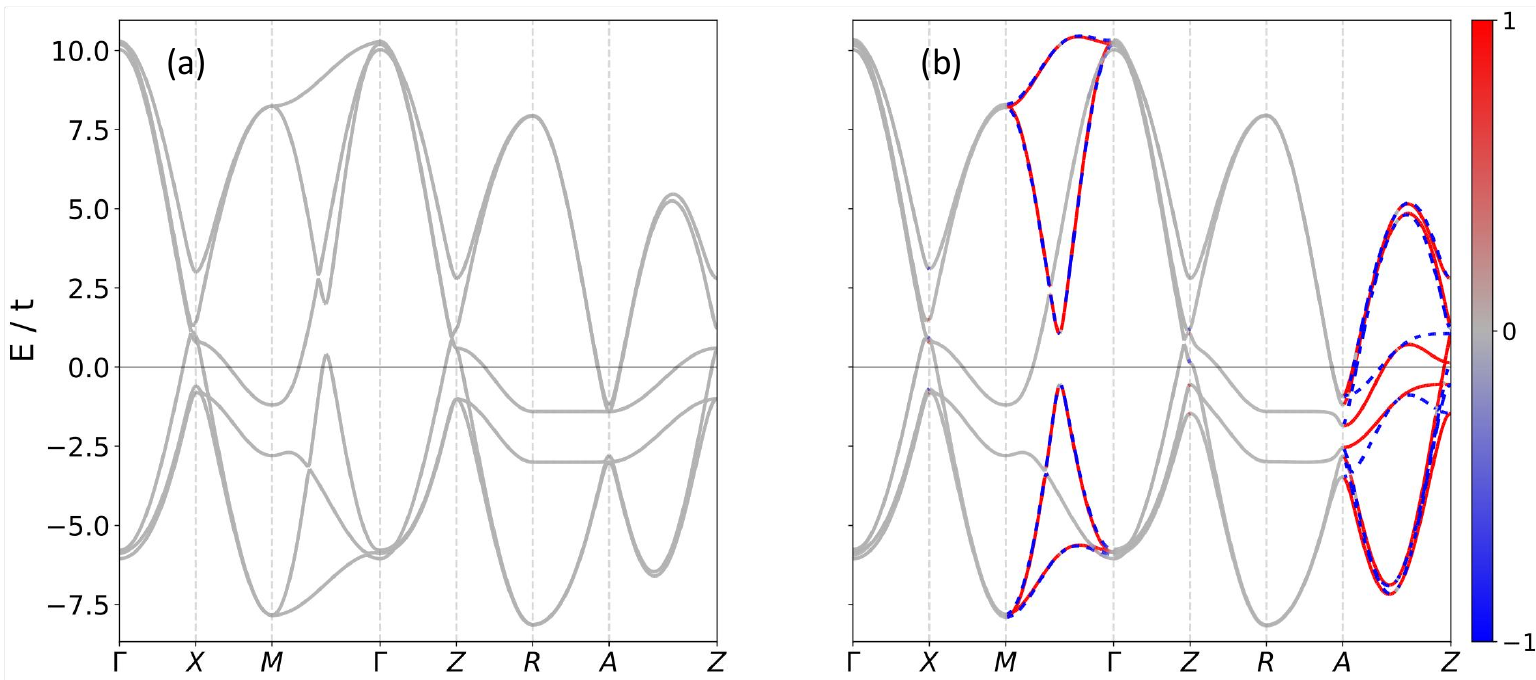}
		\caption{Energy dispersion along high symmetry line in tetragonal lattice with SOC, which contributes to the mix of the two settings of magnetization. The canting angle is set to 0. These two are without (a) and with (b) 3rd hopping. }
		\label{fig:2}
	\end{figure*}
	where the off-diagonal blocks in spin space $H_{\alpha\alpha}^{\uparrow\downarrow}$ arise from SOC and magnetization terms. Notably, the SOC and exchange coupling $J$ are site-local terms. For a fixed spin sector, the $3 \times 3$ diagonal part of the intra-sublattice hopping matrix in the orbital basis $\{d_{xy}, d_{yz}, d_{zx}\}$ is given by:
	\begin{equation}
		H_{0}^{(\alpha)} = \begin{pmatrix} \epsilon_{xy}(\boldsymbol{k}) & V_{xz}(\boldsymbol{k}) & V_{yz}(\boldsymbol{k}) \\ V_{xz}(\boldsymbol{k}) & \epsilon_{yz}(\boldsymbol{k}) & V_{xy}^{(\alpha)}(\boldsymbol{k}) \\ V_{yz}(\boldsymbol{k}) & V_{xy}^{(\alpha)}(\boldsymbol{k}) & \epsilon_{zx}(\boldsymbol{k}) \end{pmatrix},
	\end{equation}
	where the diagonal elements are:
	$\epsilon_{ij} = -2t_2(\cos k_i + \cos k_j + s\delta_{jz}\cos k_j)$.
	The off-diagonal terms are defined as:
	\begin{equation}
		\begin{aligned}
			V_{xz} &= -4t_3 \sin k_x \sin k_z, \\
			V_{yz} &= -4t_3 \sin k_y \sin k_z, \\
			V_{xy}^{(A)} &= 2t_{\text{shell}1} \cos(k_x + k_y) - 2t_{\text{shell}2} \cos(k_x - k_y), \\
			V_{xy}^{(B)} &= 2t_{\text{shell}2} \cos(k_x + k_y) - 2t_{\text{shell}1} \cos(k_x - k_y).
		\end{aligned}
	\end{equation}
	where $t_2$ and $t_3$, $t_{shell1}$, $t_{shell2}$ correspond to the next-nearest and third-nearest neighbor hoppings, respectively. $s$ is an anisotropy parameter along $z$ direction. Crucially, to satisfy the four-fold screw symmetry, the parameters $t_{shell1}$ and $t_{shell2}$ in the $V_{xy}$ term must be swapped when transforming from sublattice A to sublattice B. Consequently, the form of $H_0$ becomes sublattice-dependent (for brevity, explicit sublattice indices for $H_0$ are omitted hereafter). It is precisely this distinction in hopping parameters between the $A$ and $B$ sublattices that embodies the physical impact of the lattice symmetry specifically, the altermagnetic character. In this work, we adopt a relatively large third-nearest-neighbor hopping amplitude to accentuate and clearly observe the effects of this lattice symmetry. Specifically, the parameters used throughout the tight-binding calculations are normalized by the nearest-neighbor hopping $t_1$, setting $t_2/t_1 = 0.5$, $t_3/t_1 = 0.2$, $s = 0.2$, $J/t_1 = 0.8$, and the SOC strength $\lambda/t_1 = 0.08$. Crucially, to manifest the nature of altermagnetism, we set a relatively large disparity between the anisotropic hopping parameters, with $t_{shell1}/t_1 = 0.35$ and $t_{shell2}/t_1 = 0.12$. This significant difference explicitly controls the magnitude of the chemical environment disparity between sublattices $A$ and $B$. Consequently, the diagonal blocks $H_{AA}$ and $H_{BB}$ can be formally expressed as:
	\begin{equation}
		H_{\alpha\alpha} = \begin{pmatrix} H_0 + \frac{\lambda}{2}L_z & \frac{\lambda}{2}(L_x - iL_y) \\ \frac{\lambda}{2}(L_x + iL_y) & H_0 - \frac{\lambda}{2}L_z \end{pmatrix} + J \hat{\boldsymbol{M}_{\alpha}} \cdot \boldsymbol{\sigma},
	\end{equation}
	Note that the magnetization direction $\hat{\boldsymbol{M}}_{\alpha}$ is defined independently for the $A$ and $B$ sublattices (typically antiparallel in antiferromagnetic system). The nearest-neighbor hopping mediates the interaction between the two sublattices. Since this hopping process preserves spin, the off-diagonal blocks $H_{AB}$ and $H_{BA}$ take a block-diagonal form in spin space:
	\begin{equation}
		H_{AB} = \begin{pmatrix} H_1 & \boldsymbol{0} \\ \boldsymbol{0} & H_1 \end{pmatrix},
	\end{equation}
	where $H_1$ represents the inter-sublattice orbital hopping matrix:
	\begin{equation}
		H_{1} = \begin{pmatrix} \tau_{xy}(\boldsymbol{k}) & \Delta_{xz}(\boldsymbol{k}) & \Delta_{yz}(\boldsymbol{k}) \\ \Delta_{xz}(\boldsymbol{k}) & \tau_{yz}(\boldsymbol{k}) & \Delta_{xy}(\boldsymbol{k}) \\ \Delta_{yz}(\boldsymbol{k}) & \Delta_{xy}(\boldsymbol{k}) & \tau_{zx}(\boldsymbol{k}) \end{pmatrix},
	\end{equation}
	The matrix elements are given by $\tau_{ij}=8t_1\cos(\frac{1}{2}k_x)\cos(\frac{1}{2}k_y)\cos(\frac{1}{2}k_z)$ and $\Delta_{ij} = -8t_1^{\prime} \sin(k_i/2) \sin(k_j/2) \cos(k_l/2)$, where $\{i, j, l\}$ are permutations of the spatial indices $\{x, y, z\}$. For simplicity, we assume the two hopping coefficients are identical. The inclusion of longer-range hopping is critical for the model's symmetry. If only nearest ($t_1$) and next-nearest ($t_2$) neighbors are considered, the system retains the high symmetry of a conventional body-centered tetragonal lattice. As shown in 
	Fig. \ref{fig:2}(a), omitting the third-nearest neighbor hopping results in spin-degenerate energy bands throughout the entire Brillouin zone. However, activating the third-nearest neighbor hopping breaks this degeneracy and reveals the distinction between sublattices $A$ and $B$ Fig. \ref{fig:2}(b), which is essential for describing altermagnetism. In this regime, while the bands remain spin-degenerate along the high-symmetry boundaries of the Brillouin zone ($\Gamma-X-M$ and $\Gamma-Z-R-A$), significant spin splitting emerges along the diagonal directions ($M-\Gamma$ and $A-Z$). This momentum-dependent splitting is characteristic of $d$-wave altermagnetism. The observed band splitting pattern shows excellent agreement with first-principles calculations for $\mathrm{NiF_2}$ and $\mathrm{RuO_2}$, confirming that our tight-binding model successfully captures the essential symmetry properties of real altermagnetic materials.
	\subsection{Symmetry analysis}
\begin{figure}[!t]
		\centering\includegraphics[width=0.9\columnwidth]{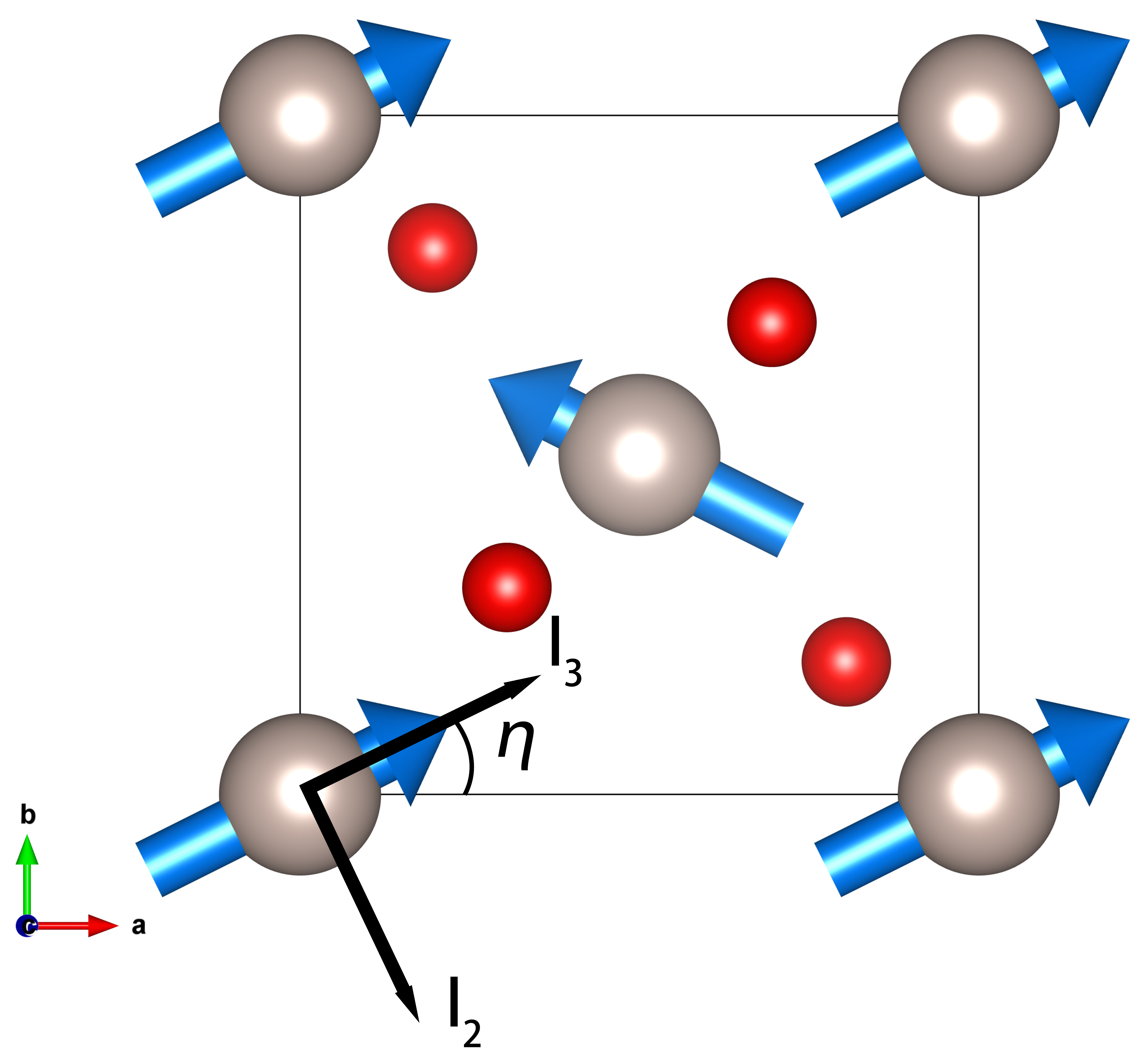}
		\caption{This figure presents a top-down view along the $z$-axis. The magnetic moments at the body-centered and corner sites are symmetric with respect to the $C_{2y}^{s}$ rotation axis. The vectors ${l}_2$ and ${l}_3$ lie within the plane, while ${l}_1$ aligns with the $z$-axis, perpendicular to the plane.}
		\label{fig:3}
	\end{figure}

	In this section, we discuss the AHE within this model. For the coplanar magnetization configuration, two distinct types of AHE exist. The first arises from the net magnetization, analogous to the AHE observed in ferromagnets. We will demonstrate that both the net magnetization induced and crystal symmetry induced contributions are primarily governed by the components of the magnetic moment projection for a given canting angle.
	In the subsequent analysis, we primarily adopt the methodology presented in Ref. \cite{PhysRevX.15.031006}. 
	Assuming the energy scale of SOC acts as a perturbation to the system, we can expand the AHC to the second order in terms of the spin orientation vectors corresponding to coplanar magnetic moments:
	\begin{equation}
		\sigma_i=\sigma_i^0+\alpha_{ij}^a\ell_j^a+\beta_{ijk}^{ab}\ell_j^a\ell_k^b+\cdots
	\end{equation}
	where $l^{a}_{i}$ denotes the spin-orbital vector representing the orientation of the spin frame. The superscript of the tensor $l$ indexes the spin space, while the subscript refers to the real space. The spin configuration of our coplanar system is illustrated in Fig. \ref{fig:3}. Its magnetic structure belongs to the magnetic space group (MSG) $Pnn'm'$ (BNS No. 58.398), which is identical to that of the collinear system with the Néel vector aligned along the $x$-axis or $y$-axis. Note that the $x$- and $y$-axes in our chosen coordinate system are interchanged relative to the standard setting adopted by the Bilbao Crystallographic Server\cite{Tasci5009,bilbao}. The MSG of the collinear system with $\eta\ne 0^\circ, 45^\circ, 90^\circ$  is $P2'/m'$ (BNS No. 10.46) with $y$-axes and $z$-axes interchanged relative to the Bilbao Crystallographic Server. The only non-vanishing AHC component of coplanar system is $\sigma_{zx}$ (equivalently denoted as $\sigma_y$) and $\sigma_{yz}$, $\sigma_{zx}$ ($\sigma_{x}$, $\sigma_{y}$) for collinear system. All three spin-frame vectors $l$ must be considered in coplanar system, and they are chosen following the recommendations in Ref. \cite{PhysRevX.15.031006}. Specifically, we align $l_3$ with one of the coplanar magnetization directions, and define $l_1$ as the cross product of the two magnetization directions to capture the relative orientation information. Consequently, the remaining vector $l_2$ is uniquely determined by the orthogonality condition. Within the framework of perturbation theory, the tensors at each order of the expansion are constrained by the symmetries present at zero SOC, namely the spin space group. As demonstrated later, it is sufficient to consider a subset of symmetry operations: $C_{2y}^SC_{2y}^L$, $C_{2z}^L$, and $C_{2y}^SC_{2x}^L$, where the superscripts $S$ and $L$ denote the operation acting on spin and orbital (spatial) space, respectively. The four-fold screw axis and the two-fold rotation axis along the diagonal have an identical effect on $\sigma$ and $l$, as will be discussed later. The transformation rules of the tensor $l$ under these three symmetry operations are summarized in Table \ref{tab:1}.
	
	\begin{table}[h]
		\centering
		
		\renewcommand{\arraystretch}{2} 
		\begin{tabular}{|c|c|c|c|}
			\hline
			& $C_{2y}^S C_{2y}^L$ & $C_{2y}^S C_{2x}^L$ & $C_{2z}^L$ \\ \hline
			$\sigma_x, \sigma_y, \sigma_z$ & $(-, +, -)$ & $(+, -, -)$ & $(-, -, +)$ \\ \hline
			$l_x^1, l_y^1, l_z^1$          & $(+, +, +)$ & $(-, -, +)$ & $(-, -, +)$ \\ \hline
			$l_x^2, l_y^2, l_z^2$          & $(+, +, +)$ & $(-, -, +)$ & $(-, -, +)$ \\ \hline
			$l_x^3, l_y^3, l_z^3$          & $(+, +, +)$ & $(-, -, +)$ & $(-, -, +)$ \\ \hline
		\end{tabular}
		\caption{The transformation rule of the AHC and the spin-orbital vector under these symmetry operation. Only component combinations that transform in the same manner as the $\sigma_{i}$ can be non-zero constituents of the $\sigma_{i}$.}
		\label{tab:1} 
	\end{table}
	
	\begin{figure*}[htbp]
		\centering\includegraphics[width=2.0\columnwidth]{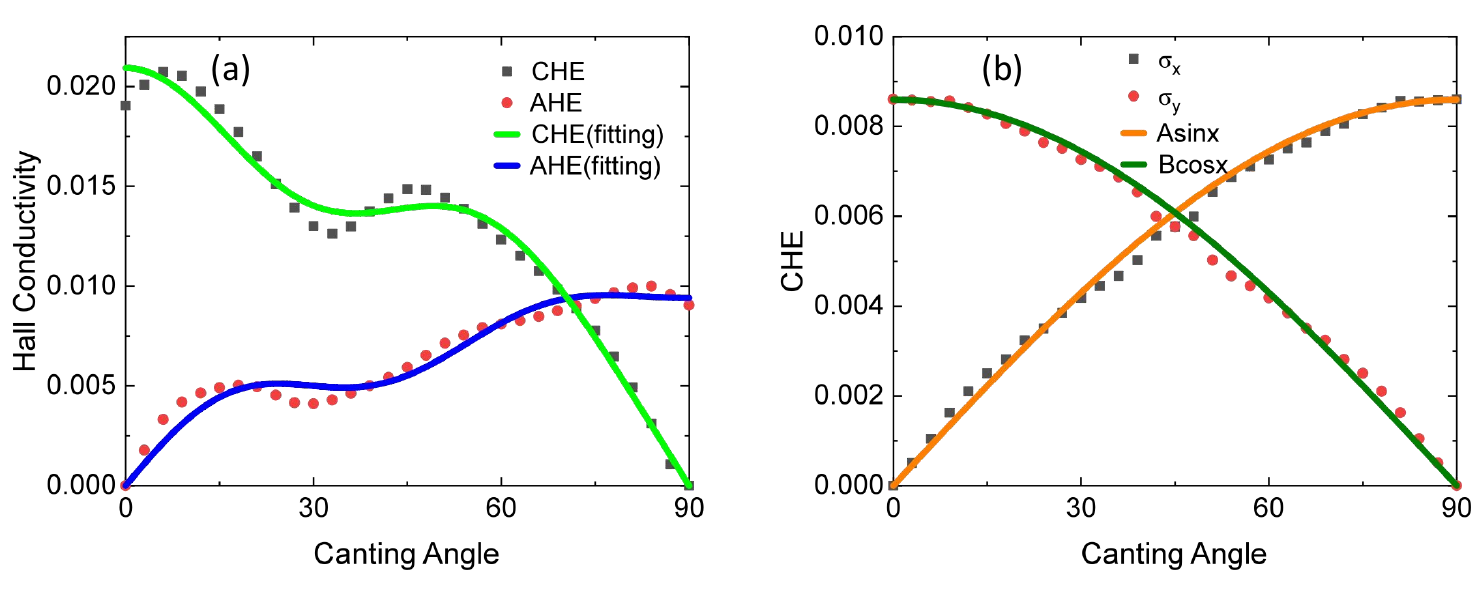}
		\caption{Calculated AHC as a function of the spin orientation angle $\eta$.
			(a) The disentanglement of AHC mechanisms in the coplanar antiferromagnetic system (Fermi energy $\mu/t = -1.25$). For visual clarity and consistency in subsequent processing, the AHC values are treated as their negatives throughout. The red circles and black squares represent the conventional magnetization-induced AHE and the crystal-symmetry-induced CHE, respectively. The solid lines indicate the theoretical trigonometric fits expanded up to the 5th order, confirming their distinct symmetry origins. The comprehensive multipole expansion basis and corresponding fitting parameters are detailed in Table\ref{tab:hall_fitting}.
			(b) The evolution of orthogonal AHC components ($\sigma_x$ and $\sigma_y$) in the collinear system (Fermi energy $\mu/t = 1$). The solid curves show the fitted results ($C=D=0.0086$). Note that the strict equivalence of $\sigma_x$ and $\sigma_y$ at specific angles is protected by the hidden $C_{110}$ rotational symmetry.}
		\label{fig:4}
	\end{figure*}
	
	Without SOC, $\sigma_i=0$, so $\sigma^{(0)}=0$. Table \ref{tab:1} shows $\sigma_{x,z}=0$ and the expansions for $\sigma_{y}$. Based on Fig. \ref{fig:3}, the vectors are:
	\begin{equation}
		\begin{aligned}
			l^{1} &= (0,0,1), \\
			l^{2} &= (\sin\eta,-\cos\eta,0), \\
			l^{3} &= (\cos\eta,\sin\eta,0).
		\end{aligned}
	\end{equation}
	We then write the non-vanishing component $\sigma_{y}$ to the first and second orders as: $\sigma_{y}^{(1)}=\alpha_{yx}^2l_{x}^{2}+\alpha_{yy}^2l_{y}^{2}+\alpha_{yx}^3l_{x}^{3}+\alpha_{yy}^3l_{y}^{3}$ and $\sigma_{y}^{(2)}=\beta_{yzx}^{12}l_{z}^{1}l_{x}^{2}+\beta_{yzy}^{12}l_{z}^{1}l_{y}^{2}+\beta_{yzx}^{13}l_{z}^{1}l_{x}^{3}+\beta_{yzy}^{13}l_{z}^{1}l_{y}^{3}$.
	
	Thus, the first- and second-order tensor of $\sigma_{y}$ can be written as:$\sigma_{y}^{(1)}=(\alpha_{yx}^3-\alpha_{yy}^2)\cos\eta+(\alpha_{yx}^2+\alpha_{yy}^3)\sin\eta$ and $\sigma_y^{(2)}=(\beta_{yzx}^{13}-\beta_{yzy}^{12})\cos\eta+(\beta_{yzx}^{12}+\beta_{yzy}^{13})\sin\eta$. We can observe that due to the presence of $l^1_{z}$, the angular dependence of the second-order terms is identical to that of the first-order terms, and similarly, the fourth-order and the sixth-order terms share the same dependence as the third-order terms. In this work, we perform the fitting for the non-coplanar system up to the seventh order as detailed in Appendix \ref{chap:expansion}.
	\begin{table}[htbp]
		\centering
		\begin{tabular}{llcc}
			\toprule
			\textbf{Order} & \textbf{Basis} & \textbf{AHE} & \textbf{CHE} \\
			\midrule
			\multirow{2}{*}{1st} 
			& $\cos(\eta)$  & $0$ & $1.94 \times 10^{-2}$ \\
			& $\sin(\eta)$  & $9.51 \times 10^{-3} $ & $0 $ \\
			\midrule
			\multirow{2}{*}{3rd} 
			& $\cos(3\eta)$  & $0$ & $-1.62 \times 10^{-3}$ \\
			& $\sin(3\eta)$  & $2.00 \times 10^{-4}$ & $0$ \\
			\midrule
			\multirow{2}{*}{5th} 
			& $\cos(5\eta)$  & $0 $ & $2.24  \times 10^{-3} $ \\
			& $\sin(5\eta)$  & $9.96 \times 10^{-4} $ & $0$ \\
			\midrule
			\multirow{2}{*}{7th} 
			& $\cos(7\eta)$  & $0 $ & $9.34 \times 10^{-4}  $ \\
			& $\sin(7\eta)$  & $9.00 \times 10^{-4} $ & $0 $ \\
			\bottomrule
		\end{tabular}
		\caption{Fitting parameters for AHE and CHE with multipole expansion basis. The detailed expansions can be found in the Appendix \ref{chap:expansion}.}
		\label{tab:hall_fitting}
	\end{table}
	
	Subsequently, we calculate the AHC of this model and compare the results with our theoretical analysis. The AHC is calculated at zero temperature by the Kubo formula:
	\begin{equation}
		\sigma_{i}=-\frac{e^2}{\hbar}\int_{BZ}\frac{d^{3}k}{(2\pi)^3}\sum_nf_n(k)\Omega_{n,i}(k)
	\end{equation}
	To disentangle the contribution induced by crystal symmetry from that induced by net magnetization, we adopt the method described in Ref. \cite{CHE}: 
	\begin{equation}
		\sigma_{i}^{\mathrm{CHE}}=[\sigma_{i}(\mathbf{n},\mathbf{m})+\sigma_{i}(\mathbf{n},-\mathbf{m})]/2,
	\end{equation}
	
	\begin{equation}
		\sigma_{i}^{\mathrm{AHE}}=[\sigma_{i}(\mathbf{n},\mathbf{m})-\sigma_{i}(\mathbf{n},-\mathbf{m})]/2.
	\end{equation}
	where $\mathbf{n}$ denotes the N\'eel vector aligned along the $x$-axis and $\mathbf{m}$ represents the net magnetization along the $y$-axis. We employ this method to disentangle the contributions of the two types of AHC in our model. Fundamentally, both mechanisms represent responses to the $l$-vector at the same order of the expansion. Specifically, the CHE corresponds to the $x$-component of the $l$-vector, which represents the magnitude of the antiferromagnetic moment along the $x$-axis, whereas the ferromagnetic AHE corresponds to the $y$-component, representing the magnitude of the ferromagnetic moment along the $y$-axis. The calculated AHC components and their corresponding trigonometric fits are presented in Fig. \ref{fig:4}(a), where the AHE exhibits a sine-like dependence and the CHE follows a cosine-like behavior. 

	For the collinear case, we calculated $\sigma_{x}$ and $\sigma_{y}$ as the N\'eel vector rotate within the $xy$ plane. The primary spin group symmetry operations are: $C_{2N}^{s}C_{2x}^{L}$ and $C_{2z}^{L}$(The $N$ direction represents a direction perpendicular to the Néel vector, which can be taken as the z-direction in our system). In the collinear case, due to the $D_{\infty}$ symmetry, the combination of $l_1$ and $l_2$ will ultimately transform to be along the $l_3$ direction\cite{PhysRevX.15.031006}. Unlike the non-collinear case, the $C_{2N}^{s}$ operation reverses the sign of all components of $l_3$ while the spatial rotation $C_{2x}^{L}$ remains identical, thus we obtain the following expressions: $\sigma_x^{(1)}=\alpha_{xy}^{3}l_y^3=\alpha_{xy}^{3}\sin\eta$ and $\sigma_y^{(1)}=\alpha_{yx}^{3}l_x^3=\alpha_{yx}^{3}\cos\eta$, with $\sigma_z=0$. The calculated results are presented in Fig. \ref{fig:4}(b). The fitting parameters for $\sigma_x$ and $\sigma_y$ are identical—specifically, 0.0086 a.u. for $\sigma_{x}$ and $\sigma_{y}$, respectively. We demonstrate below that this equivalence is exact and protected by symmetry. This behavior is attributed to the two-fold orbital rotation axis along the (110) direction  ($C_{110}^{L}$), which effectively interchanges the $x$ and $y$ coordinates. Notably, this operation belongs to the spin space group rather than the magnetic space group. Applying this symmetry to the AHC and spin vectors yields the following constraints:
	\begin{equation}
		\begin{aligned}
			\sigma_x \rightarrow \sigma_y, \quad l_y \rightarrow l_x & \implies \sigma_y^{1} = \alpha_{xy}^3 l_x \\
			& \implies \alpha_{xy}^{3}=\alpha_{yx}^{3}
		\end{aligned}
		\label{eq:c110}
	\end{equation}
	This result is in excellent agreement with our computational data. A natural question arises: why was the two-fold rotation axis along the (110) direction not considered in our previous analysis of the coplanar case? To understand this, we examine the effect of the $C^{L}_{110}$ operation. The symmetry operation $C^{S}_{2y}C^{L}_{2y}$ in the coplanar case inherently constrains each expansion term of $\sigma_x$ to be zero and thus $\alpha^{...}_{x...}$ to be zero in the specific coplanar configuration. Consider an arbitrary expansion term $\sigma^{(m)}_{y}=\alpha^{...}_{y...}l_{y}...$. Applying the $C^{L}_{110}$ operation yields $\sigma^{(m)}_{x}=\alpha^{...}_{y...}l_{x}...$ implying that for any coefficient $\alpha^{...}_{y...}$, there must exist an identical coefficient $\alpha^{...}_{x...}$. At first glance, this condition appears to require that all $\alpha^{...}_{y...}$ vanish and thus $\sigma_y$ components vanish. However, it is crucial to note that these coefficients belong to two distinct coplanar configurations, meaning that this condition constrains the expansion coefficient of $\sigma_y$ to be equal to the expansion coefficient of $\sigma_x$ in the other system. For our specific system (antiferromagnetic along the $x$-axis and ferromagnetic along the $y$-axis), the MSG operation $C_{110}$ transforms it into a configuration that is ferromagnetic along $x$ and antiferromagnetic along $y$, where $\sigma_x$ is finite. Thus, while $C_{110}$ is not a symmetry operation of the system at a specific canting angle in the presence of SOC, it acts as a bridge connecting two equivalent coplanar systems, imposing constraints on their expansion coefficients. In the collinear case, the operation $C_{110}$ acting on an antiferromagnetic moment with a tilting angle $\eta$ results in a collinear system with an angle of $(\pi/2 - \eta)$. In the small SOC limit, the corresponding SSG operation is simply $C^{L}_{110}$. Since the expansion coefficients remain invariant with respect to the tilting angle, we obtain the relation $\alpha^{3}_{xy} = \alpha^{3}_{yx}$ derived earlier. This demonstrates that when investigating the AHE in systems with varying magnetic orientations, it is essential to consider not only the local symmetry constraints at a specific angle but also the symmetries that interconnect distinct magnetic configurations and their subsequent influence on the Hall response.
	
	\subsection{Validation with real materials}
\begin{figure}[htbp]
		\centering\includegraphics[width=1\columnwidth]{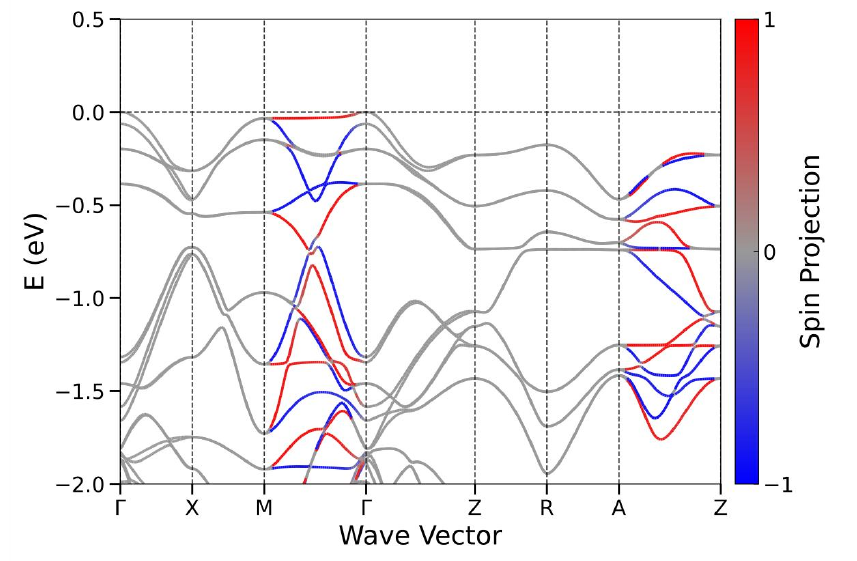}
		\caption{Calculated spin-polarized electronic band structure of $\mathrm{NiF_{2}}$ with SOC. The bands exhibit characteristic momentum-dependent spin splitting along the high-symmetry paths $M-\Gamma$ and $A-Z$, serving as a hallmark of $d$-wave altermagnetism. In contrast, the bands remain strictly spin-degenerate along the other lines due to symmetry protection. }
		\label{fig:6}
	\end{figure}

\begin{figure}[htbp]
		\centering\includegraphics[width=0.9\columnwidth]{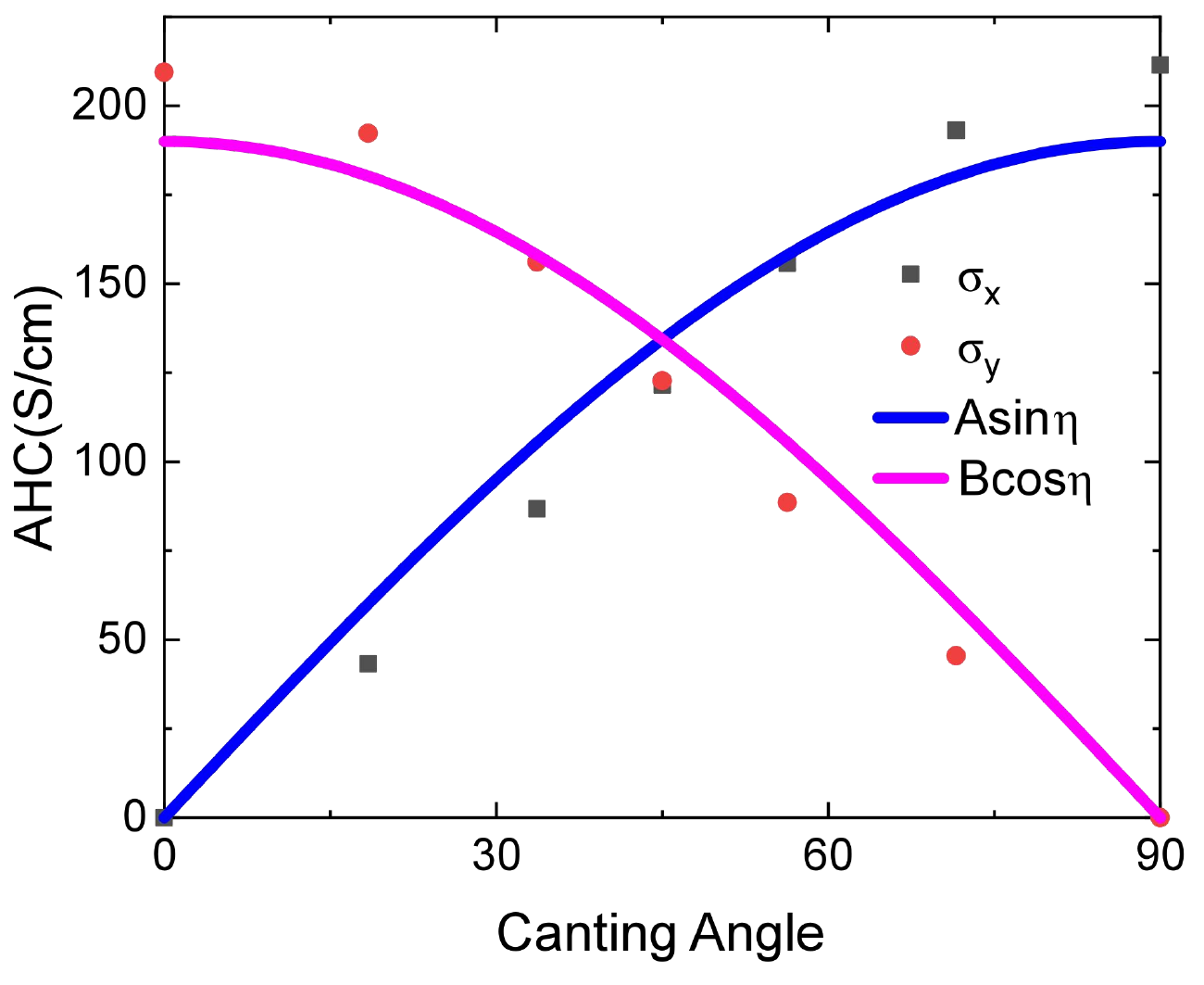}
		\caption{Calculated AHC of NiF$_2$ as a function of the canting angle $\eta$ in the collinear antiferromagnetic system (Fermi energy $\mu = -0.28$eV and $A=B=190S/cm$). Notably, the equality of $\sigma_{x}$ and $\sigma_{y}$ at $\eta=45^{\circ}$, as well as the equality of their maximum values, arises from the fact that the two collinear configurations with complementary angles are connected by the $C_{110}$ symmetry.}
		\label{fig:7}
	\end{figure}

	We calculated the spin-polarized band structure and anomalous Hall conductivity of $\mathrm{NiF_2}$\cite{NiF2}, which shares the same symmetry as our model. We performed the density functional theory (DFT) band calculation by using the Vienna Ab initio Simulation Package (VASP) \cite{kresse1996efficient}, with the generalized gradient approximation in the Perdew-Burke-Ernzerhof type as the exchange-correlation functional including SOC \cite{perdew1996generalized}.
	We set the on-site Hubbard repulsion energy $U$ to 4.0 eV for Nickel atoms\cite{NiF2}. The Brillouin zone was sampled with a $10\times10\times12$ $k$-point mesh for standard electronic structure calculations, while a significantly refined grid of $200\times200\times250$ was adopted for the AHC evaluations.
	Subsequently, we constructed the tight-binding model using Wannier90 \cite{Pizzi2020} and applied the WannierTools package \cite{WU2017} to calculate the AHC. The numerical results align well with our previous theoretical analysis. As shown in Fig. \ref{fig:6}, spin splitting is observed distinctively along the high-symmetry lines $M-\Gamma$ and $A-Z$, consistent with our earlier discussion. 
	Furthermore, calculations of the AHC for the collinear magnetic configuration confirm the role of the two-fold rotation axis $C_{110}$ along the diagonal in the realistic tetragonal system in Fig \ref{fig:7}. This axis connects two symmetry-related magnetic configurations. Consequently, the $C_{110}$ symmetry connecting two collinear cases imposes constraints on the transport coefficients, enforcing the equivalence of the expansion coefficients for $\sigma_{x}$ and $\sigma_{y}$ in the collinear system, in agreement with Eq.\ref{eq:c110}. 
	Crucially, the fitting parameters extracted from the realistic $\mathrm{NiF_2}$ calculations show excellent agreement with the symmetry analysis of the collinear case. This confirms that the specific $C_{110}$ rotation symmetry identified in our minimal three-orbital model is not an artifact of the approximation but a robust feature of the rutile structure that strictly protects the transport equivalences in real altermagnetic materials. The consistency between the DFT results and our model validates the effectiveness of our SSG analysis in predicting the evolution of topological transport properties in complex antiferromagnetic systems.

	\section{Conclusion}
	In summary, we have conducted a comprehensive theoretical investigation into the evolution of the AHC in coplanar antiferromagnetic systems. By constructing a three-orbital tight-binding model on a body-centered tetragonal lattice, we demonstrated that the inclusion of third-nearest neighbor hopping is indispensable for capturing the essential altermagnetic features—specifically, the characteristic $d$-wave spin-splitting observed in realistic materials like $\mathrm{RuO_2}$ and $\mathrm{NiF_2}$. 
	
	Leveraging a symmetry analysis based on SSG combined with a second-order perturbation expansion for SOC, we unambiguously disentangled the Hall response into two distinct origins: the conventional AHE driven by net magnetization and the CHE arising from specific crystal symmetries. We showed that these components follow precise trigonometric dependencies on the spin canting angle, a finding rigorously supported by our numerical calculations. 
	
	Most notably, we uncovered a hidden two-fold rotation symmetry, $C_{110}$, which acts as a bridge connecting distinct magnetic configurations. We clarified that this symmetry strictly protects the equivalence of orthogonal conductivity components in the collinear system. These findings not only provide a systematic framework for understanding transverse transport properties in altermagnetic materials but also offer theoretical guidance for manipulating these properties in next-generation spintronic applications.

	\addcontentsline{toc}{chapter}{Acknowledgment}
	\section*{Acknowledgment}
	This work was supported by the Science Center of the National Natural Science Foundation of China (Grant no. 12188101), the National Key R\&D Program of China (Grant nos. 2022YFA1403800, 2023YFA1607400, 2024YFA1408400), the National Natural Science Foundation of China (Grant no. 12274436). H.W. acknowledges support from the New Cornerstone Science Foundation through the XPLORER PRIZE. The governance of data was performed on the MatElab platform, developed by the Condensed Matter Physics Data Center of Chinese Academy of Sciences. The AI-driven experiments, simulations, and model training were performed on the robotic AI-Scientist platform of the Chinese Academy of Sciences.
	
	\appendix
	
	\section{Expansion of CHE and AHE}
	\label{chap:expansion}
	Referring to Table \ref{tab:1}, we can obtain the non-vanishing terms of the third-order expansion:
	\begin{equation}
		\begin{aligned}
			\sigma_{y}^{(3)} =& A\sin\eta + B\cos\eta + C\sin^3\eta + D\cos^3\eta \\
			&+ E\sin^2\eta\cos\eta + F\sin\eta\cos^2\eta
		\end{aligned}
	\end{equation}
	where the effective coefficients are:
	
	\begin{equation}
		\begin{aligned}
			A = &\quad 3\gamma^{112}_{zzx} + 3\gamma^{113}_{zzy} \\
			B = &-3\gamma^{112}_{zzy} + 3\gamma^{113}_{zzx}  \\
			C = &\quad \gamma^{222}_{xxx} + 3\gamma^{223}_{xxy}  + 3\gamma^{233}_{xyy} + \gamma^{333}_{yyy} \\
			D = &-\gamma^{222}_{yyy} + 3\gamma^{223}_{yyx}  - 3\gamma^{233}_{yxx} + \gamma^{333}_{xxx}\\
			E =& -3\gamma^{222}_{xxy} + (3\gamma^{223}_{xxx} - 6\gamma^{223}_{xyy} ) \\
			&+ (6\gamma^{332}_{xyx} - 3\gamma^{332}_{yyy}) + 3\gamma^{333}_{xyy}  \\
			F =& \quad 3\gamma^{222}_{xyy} + (-6\gamma^{223}_{xyx} + 3\gamma^{223}_{yyy}) \\
			&+(3\gamma^{233}_{xxx} - 6\gamma^{233}_{yxy}) 
			+3\gamma^{333}_{xxy} 
		\end{aligned}
	\end{equation}
	It can be expressed in this basis:
	\begin{equation}
		\begin{aligned}
			\sigma_{y}^{(3)} =& a\sin\eta + b\cos\eta + c\sin(3\eta) + d\cos(3\eta) 
		\end{aligned}
	\end{equation}
	The expansions in Table \ref{tab:hall_fitting} can be expressed as:
	\begin{equation}
		\begin{aligned}
			\sigma_{\text{CHE}}(\eta) =& A \cos\eta +B \cos(3\eta) + C \cos(5\eta) + D \cos(7\eta), \\
			\sigma_{\text{AHE}}(\eta) =& a \sin\eta + b \sin(3\eta) +c \sin(5\eta) + d \sin(7\eta).
		\end{aligned}
	\end{equation}
	with:
	\begin{equation}
		\begin{aligned}
			A = &\quad 1.94 \times 10^{-2}\\
			B = &- 1.62 \times 10^{-3}  \\
			C = &\quad 2.24 \times 10^{-3} \\
			D = &-9.34 \times 10^{-4}\\
			a =& \quad 9.51 \times 10^{-3} \\
			b =& \quad 2.00 \times 10^{-4} \\
			c =& \quad 9.96 \times 10^{-4} \\
			d =& \quad 9.00 \times 10^{-4}
		\end{aligned}
	\end{equation}

	\addcontentsline{toc}{chapter}{References}
	\bibliographystyle{iopart-num} 
	
	
	\bibliography{ref}
	
	
\end{document}